\documentstyle[psfig,preprint,aps]{revtex}
\newcommand {\ignore}[1]{}
\def\lsim{\:\raisebox{-0.5ex}{$\stackrel{\textstyle<}{\sim}$}\:}
\def\gsim{\:\raisebox{-0.5ex}{$\stackrel{\textstyle>}{\sim}$}\:}
\newcommand{\Frac}[2]{\frac{\displaystyle #1}{\displaystyle #2}}

\def\21{$SU(2) \ot U(1)$}
\def\321{$SU(3) \ot SU(2) \ot U(1)$}
\def\ne{\hbox{$\nu_e$ }}
\def\nm{\hbox{$\nu_\mu$ }}
\def\nt{\hbox{$\nu_\tau$ }} 
\def\ns{\hbox{$\nu_{s}$ }}

\def\apj#1#2#3{          { Astrophys. J. }{\bf #1}, #3 (19#2)}

\def\ib#1#2#3{           { ibid. }{\bf #1}, #3 (19#2)}

\def\np#1#2#3{           { Nucl. Phys. }{\bf #1}, #3 (19#2)}
\def\pl#1#2#3{           { Phys. Lett. }{\bf #1}, #3 (19#2)}
\def\pr#1#2#3{           { Phys. Rev. }{\bf #1}, #3 (19#2)}

\def\prl#1#2#3{          { Phys. Rev. Lett. }{\bf #1}, #3 (19#2)}

\def\zp#1#2#3{           { Zeit. fur Physik }{\bf #1}, #3 (19#2)}

\def\n.c.#1#2#3{         { Nuovo Cim. }{\bf #1}, #3 (19#2)}
\def\r.n.c.#1#2#3{       { Riv. del Nuovo Cim. }{\bf #1}, #3 (19#2)}

\begin{document}
\draft
\preprint{\vbox{
\hbox{hep-ph/9807305}
\hbox{FTUV/98-56, IFIC/98-57}
} }
\renewcommand{\baselinestretch}{1.1}
\title{Active-active and active-sterile neutrino  
oscillation solutions to the atmospheric neutrino anomaly}

\author{M.\ C.\ Gonzalez-Garcia$^{1}$
\footnote{E-mail concha@evalvx.ific.uv.es}, 
H.\ Nunokawa$^{2}$
\footnote{E-mail nunokawa@ifi.unicamp.br}, 
O.\ L.\ G.\ Peres$^{1}$
\footnote{E-mail operes@flamenco.ific.uv.es}
and \\ 
J.\ W.\ F.\ Valle$^1$ 
\footnote{E-mail valle@flamenco.ific.uv.es}}
\address{$^1$ \it Instituto de F\'{\i}sica Corpuscular - C.S.I.C.\\
Departament de F\'{\i}sica Te\`orica, Universitat de Val\`encia\\
46100 Burjassot, Val\`encia, Spain \\
http://neutrinos.uv.es
}
\address{${}^2$Instituto de F\'{\i}sica Gleb Wataghin, 
         Universidade Estadual de Campinas\\
          13083-970 Campinas, S\~ao Paulo, Brazil}
\maketitle

\renewcommand{\baselinestretch}{.8}
\begin{abstract}
We perform a fit to the full data set corresponding to $33.0$ kt-yr of
data of the Super-Kamiokande experiment as well as to all other
experiments in order to compare the two most likely solutions to the
atmospheric neutrino anomaly in terms of oscillations in the $\nu_\mu
\to \nu_\tau$ and $\nu_\mu \to \nu_s$ channels. Using state-of-the-art
atmospheric neutrino fluxes we have determined the allowed regions of
oscillation parameters for both channels. We find that the $\Delta
m^2$ values for the active-sterile oscillations (both for positive and
negative $\Delta m^2$) are higher than for the $\nu_\mu \to \nu_\tau$
case, and that the increased Super-Kamiokande sample slightly favours
$\nu_\mu \to \nu_\tau$ oscillations over oscillations into a sterile
species $\nu_s$, $\nu_\mu \to \nu_s$, and disfavours $\nu_\mu \to
\nu_e$. We also give the zenith angle distributions predicted for 
the best fit points in each of the possible oscillation channels.
Finally we compare our determinations of the atmospheric neutrino
oscillation parameters with the expected sensitivities of future
long-baseline experiments K2K, MINOS, ICARUS, OPERA and NOE.

\pacs{14.60.Pq, 13.15.+g, 95.85.Ry}

\end{abstract}

\section{Introduction}

Atmospheric showers are initiated when primary cosmic rays hit the
Earth's atmosphere. Secondary mesons produced in this collision,
mostly pions and kaons, decay and give rise to electron and muon
neutrino and anti-neutrinos fluxes \cite{review}.  There has been a
long-standing anomaly between the predicted and observed $\nu_\mu$
$/\nu_e$ ratio of the atmospheric neutrino fluxes
\cite{atmexp+}. Although the absolute individual $\nu_\mu$ or $\nu_e$
fluxes are only known to within $30\%$ accuracy, different authors
agree that the $\nu_\mu$ $/\nu_e$ ratio is accurate up to a $5\%$
precision. In this resides our confidence on the atmospheric neutrino
anomaly (ANA), now strengthened by the high statistics sample
collected at the Super-Kamiokande experiment \cite{superkam}.  This
experiment has marked a turning point in the significance of the
ANA. The most likely solution of the ANA involves neutrino
oscillations \cite{osciat}.  
In principle we can invoke various neutrino oscillation
channels, involving the conversion of \nm into either \ne or \nt
(active-active transitions) or the oscillation of \nm into a sterile
neutrino \ns (active-sterile transitions). Previously we have reported
an exhaustive analysis of all available experimental atmospheric
neutrino data for $\nu_\mu\to \nu_e$ and for $\nu_\mu\to \nu_{\tau}$
channels ~\cite{ourwork} but we have not discussed the sterile
neutrino case. This is especially well-motivated theoretically, since
it constitutes one of the simplest ways to reconcile
\cite{BEYONDTHEDESERT,PTV,all} the ANA with other puzzles in the
neutrino sector such as the solar neutrino problem \cite{SNP,bahcall} as well
as the LSND result \cite{lsnd} and the possible need for a few eV mass
neutrino as the hot dark matter in the Universe \cite{SS1,PHKC}.
Although stringent limits on the existence of sterile states with
large mixing to standard neutrinos have been obtained from
cosmological Big-Bang Nucleosynthesis considerations
\cite{chicago}, more conservative estimations claim that the relevant
effective number of light neutrino degrees of freedom may still allow
or even exceed $N_\nu=4$ ~\cite{sarkar}, thus not precluding scenarios
such as given in \cite{PTV}. Moreover, nucleosynthesis bounds might be
evaded due to the possible suppression of active-sterile neutrino
conversions in the early universe due to the presence of a lepton
asymmetry that can be generated by the oscillations themselves
~\cite{foot1}.

The main aim of the present paper is to compare the $\nu_\mu \to
\nu_\tau$ and the $\nu_\mu \to \nu_{s}$ transitions using the the 
new sample corresponding to 535 days of the Super-Kamiokande
data. Our present analysis uses the latest improved calculations of
the atmospheric neutrino fluxes as a function of zenith angle
\cite{agrawal}, including the muon polarization effect ~\cite{volkova}
and taking into account a variable neutrino production point
~\cite{pathlenght}. 
We determine the allowed regions for active-sterile 
oscillation parameters for the two possible cases $\Delta m^2 >0$ and
$\Delta m^2 <0$. We first analyze the data using only the
Super-Kamiokande results up to 535 days. Then we also perform a
global analysis including the results of the Soudan2, IMB, Frejus,
Nusex and Kamiokande experiments. Instead of using directly the
double-ratio our analysis relies on the separate use of the electron
and $\mu$ type event numbers and zenith angle distributions, with
their error correlations. Following ref.~\cite{ourwork} we explicitly
check the agreement of our theoretical predictions with the
experimental Monte Carlo, rendering reliability on our method.

Our main result is that the $\Delta m^2$ values for the active-sterile
oscillations (both for positive and negative $\Delta m^2$) are higher
than for the $\nu_\mu \to \nu_\tau$ case with maximal or nearly
maximal mixing. We find that the increased Super-Kamiokande sample
slightly favours $\nu_\mu \to \nu_\tau$ oscillations in comparison
with $\nu_\mu \to \nu_s$, and strongly disfavours $\nu_\mu \to \nu_e$.
Notice that this in contrast with our previous results in 
Ref. \cite{ourwork} were the smaller Super-Kamiokande sample did not
have enought statistical weight to overcome the best fit of the 
$\nu_\mu \to \nu_e$ channel to the Kamiokande multi-GeV data.
We find that, even though the high Super-Kamiokande statistics plays a
major role in the determination of the final allowed region, there is
still some appreciable weight carried by the other experiments.  In
the global fit case the preference for the $\nu_\mu \to \nu_\tau$
channel over the $\nu_\mu \to \nu_s$ or $\nu_\mu \to \nu_e$ channels
is marginal, in clear contrast to the situation where only the
Super-Kamiokande data is taken into account.  Finally we compare our
results of the atmospheric neutrino oscillation parameters with the
accelerator neutrino oscillation searches such as CDHSW \cite{CDHSW},
CHORUS-NOMAD \cite{dilella}, as well as searches at reactor neutrino
experiments \cite{krasnoyarsk,bugey,chooz}. We re-confirm our previous
conclusions \cite{ourwork} that the recent data obtained at
long-baseline reactor experiment CHOOZ ~\cite{chooz} rules out at the
90 \% CL the \nm $\to$ \ne channel as a solution to the ANA.  We also
compare the results of our fit with the expected sensitivities of
future long-baseline experiments such as K2K\cite{chiaki},
MINOS\cite{minos}, ICARUS\cite{icarus} and NOE\cite{noe}.

\section{Atmospheric Neutrino Oscillation Probabilities}

Here we determine the expected neutrino event number both in the
absence and the presence of oscillations.  First we compute the
expected number of $\mu$-like and $e$-like events, $N_\alpha$, $\alpha
= \mu, e$ for each experiment
\begin{equation}
N_\mu= N_{\mu\mu} +\
 N_{e\mu} \; ,  \;\;\;\;\;\
N_e= N_{ee} +  N_{\mu e} \; ,
\label{eventsnumber}
\end{equation}
where
\begin{equation}
N_{\alpha\beta} = n_t T
\int
\frac{d^2\Phi_\alpha}{dE_\nu d(\cos\theta_\nu)} 
\kappa_\alpha(h,\cos\theta_\nu,E_\nu)
P_{\alpha\beta}
\frac{d\sigma}{dE_\beta}\varepsilon(E_\beta)
dE_\nu dE_\beta d(\cos\theta_\nu)dh\; .
\label{event0}
\end{equation}
and $P_{\alpha\beta}$ is the oscillation probability of $\nu_\alpha \to
\nu_\beta$ for given values of $E_{\nu}, \cos\theta_\nu$ and $h$,
i.e., $P_{\alpha\beta} \equiv P(\nu_\alpha \to \nu_\beta; E_\nu,
\cos\theta_\nu, h ) $.  In the case of no oscillations, the only
non-zero elements are the diagonal ones, i.e. $P_{\alpha\alpha}=1$ for
all $\alpha$.

Here $n_t$ is the number of targets, $T$ is the experiment's running
time, $E_\nu$ is the neutrino energy and $\Phi_\alpha$ is the flux of
atmospheric neutrinos of type $\alpha=\mu ,e$; $E_\beta$ is the final
charged lepton energy and $\varepsilon(E_\beta)$ is the detection
efficiency for such charged lepton; $\sigma$ is the neutrino-nucleon
interaction cross section, and $\theta_\nu$ is
the angle between the vertical direction and the incoming neutrinos
($\cos\theta_\nu$=1 corresponds to the down-coming neutrinos).  In
Eq.~(\ref{event0}), $h$ is the slant distance from the production
point to the sea level for $\alpha$-type neutrinos with energy $E_\nu$
and zenith angle $\theta_\nu$. Finally, $\kappa_\alpha$ is the slant
distance distribution which is normalized to one \cite{pathlenght}.

The neutrino fluxes, in particular in the sub-GeV range, depend on the
solar activity.  In order to take this fact into account we use in
Eq.~(\ref{event0}) a linear combination of atmospheric neutrino fluxes
$\Phi_\alpha^{max}$ and $\Phi_\alpha^{min}$ which correspond to the
most active Sun (solar maximum) and quiet Sun (solar minimum),
respectively, with different weights, depending on the running period
of each experiment \cite{ourwork}.
  
For definiteness we assume a two-flavour oscillation scenario, in which
the $\nu_\mu$ oscillates into another flavour either $\nu_\mu \to
\nu_e$ , $\nu_\mu \to \nu_s$ or $\nu_\mu \to
\nu_\tau$. The Schr\"oedinger evolution equation of the $\nu_\mu -\nu_X$ 
(where $X=e,\tau $ or $s$ sterile) system in the matter background for
{\sl neutrinos } is given by \cite{wolf}
\begin{eqnarray}
i{\mbox{d} \over \mbox{d}t}\left(\matrix{
\nu_\mu \cr\ \nu_X\cr }\right) & = & 
 \left(\matrix{
 {H}_{\mu}
& {H}_{\mu X} \cr
 {H}_{\mu X} 
& {H}_X \cr}
\right)
\left(\matrix{
\nu_\mu \cr\ \nu_X \cr}\right) \,\,, \\
  H_\mu & \! = &  \! 
 V_\mu + \frac{\Delta m^2}{4E_\nu} \cos2 \theta_{\mu X}\,, \,\,\,\,\,\,\,\,\,
H_X \!= V_X -  \frac{\Delta m^2}{4E_\nu} \cos2 \theta_{\mu X}, \nonumber \\
H_{\mu X}& \!= &  - \frac{\Delta m^2}{4E_\nu} \sin2 \theta_{\mu X}, \\
\label{evolution1}
\end{eqnarray}
where 
\begin{eqnarray}
\label{potential}
V_\tau=V_\mu & = &\frac{\sqrt{2}G_F \rho}{M} (-\frac{1}{2}Y_n)\,, 
\\
V_s &= & 0\,,\\
V_e & = & \frac{\sqrt{2}G_F \rho}{M} ( Y_e - \frac{1}{2}Y_n)
\\
\nonumber
\end{eqnarray}
Here $G_F$ is the Fermi constant, $\rho$ is the matter density at the
Earth, $M$ is the nucleon mass, and $Y_e$ ($Y_n$) is the electron
(neutron) fraction. We define $\Delta m^2=m_2^2-m_1^2$ in such a way
that if $\Delta m^2>0 \: (\Delta m^2<0)$ the neutrino with largest
muon-like component is heavier (lighter) than the one with largest
X-like component. For anti-neutrinos the signs of potentials $V_X$
should be reversed. We have used the approximate analytic expression
for the matter density profile in the Earth obtained in
ref. \cite{lisi}. In order to obtain the oscillation probabilities
$P_{\alpha\beta}$ we have made a numerical integration of the
evolution equation. The probabilities for neutrinos and anti-neutrinos
are different because the reversal of sign of matter potential. Notice
that for the $\nu_\mu\to\nu_\tau$ case there is no matter effect while
for the $\nu_\mu\to\nu_s$ case we have two possibilities depending on
the sign of $\Delta m^2$.  For $\Delta m^2 > 0$ the matter effects
enhance {\sl neutrino} oscillations while depress {\sl anti-neutrino}
oscillations, whereas for the other sign ($\Delta m^2<0$) the opposite
holds. The same occurs also for $\nu_\mu\to\nu_e$.  Although in the
latter case one can also have two possible signs, we have chosen the
most usually assumed case where the muon neutrino is heavier than the
electron neutrino, as it is theoretically more appealing. Notice also
that, as seen later, the allowed region for this sign is larger than
for the opposite, giving the most conservative scenario when comparing
with the present limits from CHOOZ.

\section{Atmospheric Neutrino Data Fits} 

Here we describe our fit method to determine the atmospheric
oscillation parameters for the various possible oscillation channels,
including matter effects for both $\nu_\mu \to \nu_e$ and $\nu_\mu \to
\nu_s$ channels. In the latter case we consider both $\Delta m^2$ signs.
Some results for the $\nu_\mu \to \nu_\tau$ and $\nu_\mu \to \nu_s$
channels have already been presented in ref.  ~\cite{yasuda1}.
However our approach is more complete and systematic, complementing
that of ~\cite{yasuda1} in many ways. For example ref. ~\cite{yasuda1}
uses the double ratio of experimental-to-expected ratio of muon-like
to electron-like events $R_{\mu/e}/R^{MC}_{\mu/e}$ and the
upward-going/down-going muon ratio~\cite{pakvasa}. It is well known
that the double-ratio is not well-suited from a statistical point of
view due to its non-Gaussian character \cite{fogli2}. Although the
Super-Kamiokande statistics is better, we prefer to rely on the
separate use of the event numbers paying attention to the correlations
between the muon predictions and electron predictions.  Moreover,
following ref.~\cite{ourwork} we explicitly verify in our present
reanalysis the agreement of our predictions with the experimental
Monte Carlo predictions, leading to a good confidence in the
reliability of our results. We also use the atmospheric neutrino flux
calculation of ref. ~\cite{agrawal} instead of the Honda et al fluxes
ref. ~\cite{honda}. Last but not least we perform a global analysis of
the data, instead of focussing only on Super-Kamiokande data, admitting
however the main role played by the latter experiment. In this case we
take especial care in implementing the experimental detection
efficiencies \cite{inoue}, as in ref.~\cite{ourwork}.

The steps required in order to generate the allowed regions of
oscillation parameters were given in ref. \cite{ourwork}.  As
already mentioned we focus on the simplest interpretation of the ANA
in terms of the neutrino oscillation hypothesis.  For definiteness we
assume a two-flavour oscillation scenario, in which the $\nu_\mu$
oscillates into another flavour either $\nu_\mu \to \nu_e$ , $\nu_\mu
\to \nu_s$ or $\nu_\mu \to
\nu_\tau$.

As already mentioned, when combining the results of the experiments we
do not make use of the double ratio, $R_{\mu/e}/R^{MC}_{\mu/e}$, but
instead we treat the $e$ and $\mu$-like data separately, taking into
account carefully the correlation of errors. Following
ref. \cite{ourwork,fogli2} we define the $\chi^2$ as
\begin{equation}
\chi^2 \equiv \sum_{I,J}
(N_I^{data}-N_I^{theory}) \cdot 
(\sigma_{data}^2 + \sigma_{theory}^2 )_{IJ}^{-1}\cdot 
(N_J^{data}-N_J^{theory}),
\label{chi2}
\end{equation}
where $I$ and $J$ stand for any combination of the experimental data
set and event-type considered, i.e, $I = (A, \alpha)$ and $J = (B,
\beta)$ where, $A,B$ stands for Fr\'ejus, Kamiokande sub-GeV, IMB,... and
$\alpha, \beta = e,\mu$.  In Eq.~(\ref{chi2}) $N_I^{theory}$ is the
predicted number of events calculated from Eq.~(\ref{eventsnumber})
whereas $N_I^{data}$ is the number of observed events.  In
Eq.~(\ref{chi2}) $\sigma_{data}^2$ and $\sigma_{theory}^2$ are the
error matrices containing the experimental and theoretical errors
respectively. They can be written as
\begin{equation}
\sigma_{IJ}^2 \equiv \sigma_\alpha(A)\, \rho_{\alpha \beta} (A,B)\,
\sigma_\beta(B),
\end{equation}
where $\rho_{\alpha \beta} (A,B)$ stands for the correlation between
the $\alpha$-like events in the $A$-type experiment and $\beta$-like
events in $B$-type experiment, whereas $\sigma_\alpha(A)$ and
$\sigma_\beta(B)$ are the errors for the number of $\alpha$ and
$\beta$-like events in $A$ and $B$ experiments, respectively. The
dimension of the error matrix varies depending on the combination of
experiments included in the analysis.  

We compute $\rho_{\alpha \beta} (A,B)$ as in ref. \cite{fogli2}.  A
detailed discussion of the errors and correlations used in our
analysis can be found in Ref.~\cite{ourwork}.  In our present
analysis, we have conservatively ascribed a 30\% uncertainty to the
absolute neutrino flux, in order to generously account for the spread
of predictions in different neutrino flux calculations.  Next we
minimize the $\chi^2$ function in Eq.~(\ref{chi2}) and determine the
allowed region in the $\sin^22\theta-\Delta m^2$ plane, for a given
confidence level, defined as,
\begin{equation}
\chi^2 \equiv \chi_{min}^2  + 4.61\ (9.21)\   \  
\  \mbox{for}\  \ 90\  (99) \% \ \  \mbox{C.L.}
\label{chimin}
\end{equation}

There are some minor changes with respect to the assumed errors quoted
in ref.~\cite{ourwork}, mainly for the Super-Kamiokande experiment.
These are the following \cite{Kearns}:
\begin{itemize}
\item
The error in the $\mu/e$ ratio, arising from the error in the charged
current cross sections. This is estimated to be $4.3\%$.
\item
The error in the electron event number arising from the uncertainty in
the neutral current cross section, which is estimated to be $3.0\%$
($4.1\%$ ) for sub-GeV (multi-GeV) events, respectively.
\item
The electron versus muon mis-identification error for Super Kamiokande
multi-GeV events. This is estimated to be $1.5\%$.
\end{itemize}

\begin{figure}
\centerline{\protect\hbox{\psfig{file=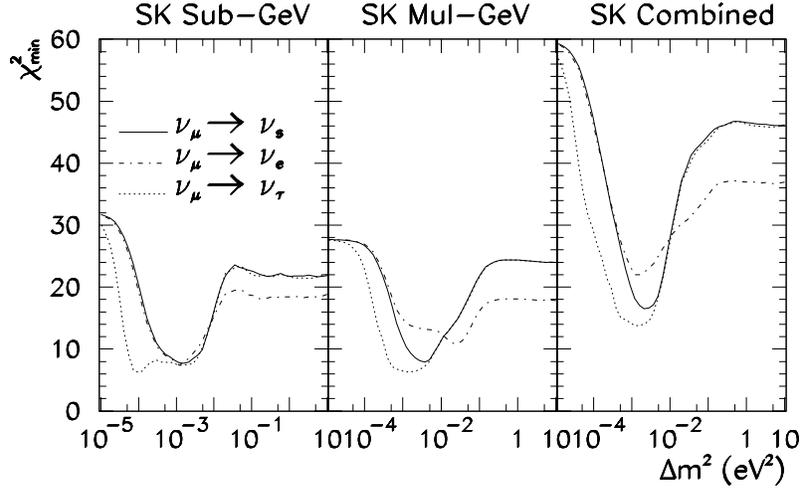,width=0.75\textwidth}}}
\vglue 1cm
\caption{$\chi^2_{min}$ for fixed $\Delta m^2$ versus $\Delta m^2$ for
each oscillation channel for Super-Kamiokande sub-GeV and multi-GeV data, 
and for the combined sample. Since the minimum is always obtained
close to maximum mixing the curves for $\nu_\mu\rightarrow \nu_s$ for both
signs of $\Delta m^2$ coincide.}
\label{chimin1} 
\end{figure}

In Fig.~\ref{chimin1} we plot the minimum value attained by the
$\chi^2$ function for fixed $\Delta m^2$ as $\sin^2 2\theta$ is varied
freely, as a function of $\Delta m^2$.  Notice that for large $\Delta
m^2 \gsim 0.1$ eV$^2$, the $\chi^2$ is nearly constant.  This happens
because in this limit the contribution of the matter potential in
Eq~(\ref{evolution1}) can be neglected with respect to the $\Delta
m^2$ term, so that the matter effect disappears and moreover, the
oscillation effect is averaged out. In fact one can see that in this
range we obtain nearly the same $\chi^2$ for the $\nu_\mu \to
\nu_\tau$ and $\nu_\mu \to \nu_{s}$ cases. For very small 
$\Delta m^2 \lsim 10^{-4}$ eV$^2$, the situation is opposite, namely
the matter term dominates and we obtain a better fit for the $\nu_\mu
\to \nu_\tau$ channel, as can be seen by comparing the $\nu_\mu \to
\nu_{\tau}$ curve of the Super-Kamiokande sub-GeV data (dotted curve
in the left panel of Fig.~\ref{chimin1}) with the solid ($\nu_\mu \to
\nu_s$) and dashed ($\nu_\mu \to \nu_e$) curves in the left panel of
Fig.~\ref{chimin1}). For extremely small $\Delta m^2 \lsim 10^{-4}$
eV$^2$, values $\chi^2$ is quite large and approaches a constant,
independent of oscillation channel, as in the no-oscillation case.
Since the average energy of Super-Kamiokande multi-GeV data is higher
than the sub-GeV one, we find that the limiting $\Delta m^2$ value
below which $\chi^2$ approaches a constant is higher, as seen in the
middle panel. Finally, the right panel in Fig.~\ref{chimin1} is
obtained by combining sub and multi-GeV data.

A last point worth commenting is that for the $\nu_\mu \to \nu_{\tau}$
case in the sub-GeV sample there are two almost degenerate values of
$\Delta m^2$ for which $\chi^2$ attains a minimum. From
Fig.~\ref{chimin1} one sees that the corresponding oscillation
parameter values are $\Delta m^2=1.1 \times 10^{-4}$ eV$^2$ and $2.4
\times 10^{-3} $eV$^2$, both with maximal mixing.  For the multi-GeV 
case there is just one minimum at $1.4 \times 10^{-3} $eV$^2$. Finally
in the third panel in Fig.~\ref{chimin1} we can see that by
combining the Super-Kamiokande sub-GeV and multi-GeV data we have a
unique minimum at $1.4 \times 10^{-3} $eV$^2$.

\section{Results for the Oscillation Parameters}
 
\begin{figure}
\centerline{\protect\hbox{\psfig{file=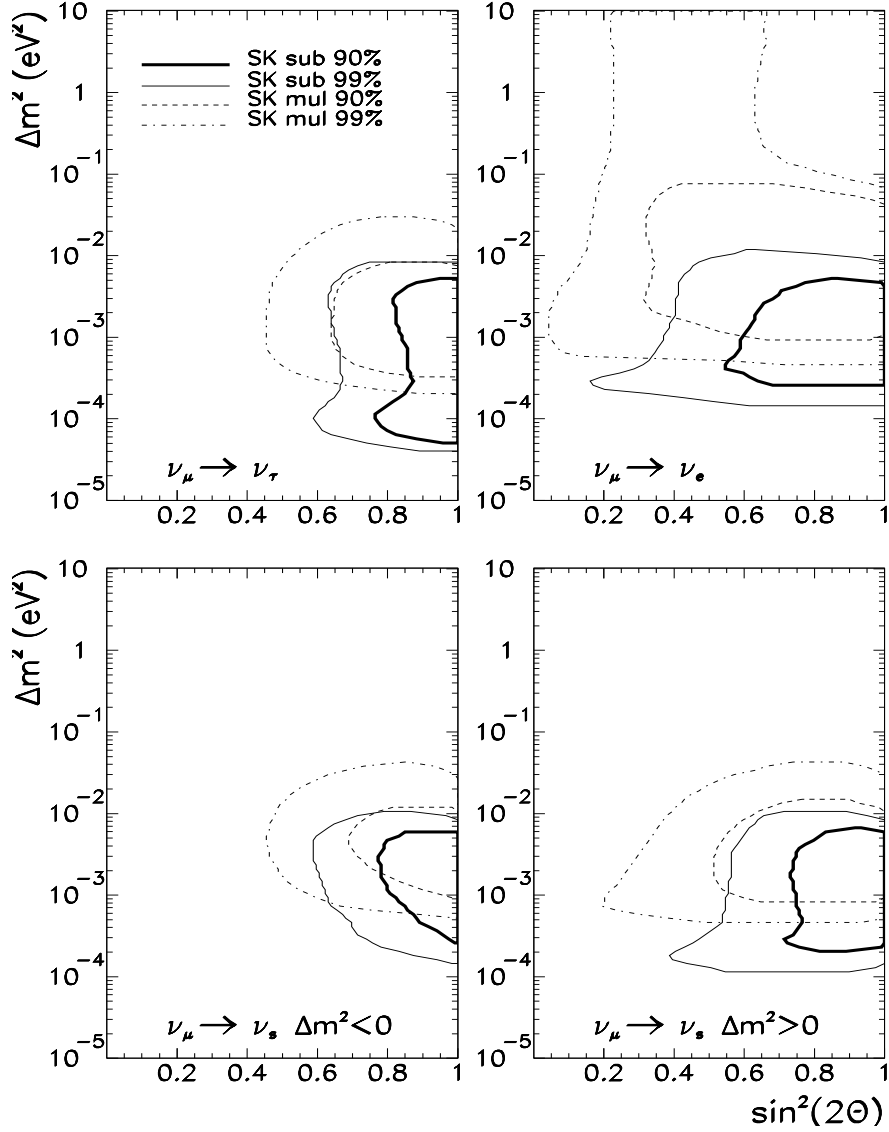,width=0.8\textwidth,height=0.7\textheight}}}
\vglue 1cm
\caption{
Allowed regions of oscillation parameters for Super-Kamiokande for 
the different oscillation channels as labeled in the figure. 
In each panel, we show the allowed regions for the sub-GeV data  
at 90 (thick solid line) and 99 \% CL  (thin solid line)
and the multi-GeV data at 90 (dashed line) and 99 \% CL  (dot-dashed line).}
\label{mutausk1} 
\end{figure}

The results of our $\chi^2$ fit of the Super-Kamiokande sub-GeV and
multi-GeV atmospheric neutrino data are given in Fig.~\ref{mutausk1}.
In this figure we give the allowed region of oscillation parameters at 90
and 99 \% CL.  The upper left, upper right, lower left and lower right
panels show respectively the $\nu_\mu \to \nu_\tau$, $\nu_\mu \to
\nu_e$ ($\Delta m^2 >0$), $\nu_\mu \to \nu_s$ ($\Delta m^2<0$) and 
the $\nu_\mu \to \nu_s$ ($\Delta m^2>0$). The thick solid (thin solid)
curves show the sub-GeV region at $90$ ($99$) \% CL regions and the
dashed (dot-dashed) curves the multi-GeV region at $90$ ($99$) \% CL
regions.

One can notice that the matter effects are similar for the upper right
and lower right panels because matter effects enhance the oscillations
for {\sl neutrinos} in both cases.  In contrast, in the case of
$\nu_\mu \to \nu_s$ with $\Delta m^2<0$ the enhancement occurs only
for {\sl anti-neutrinos} while in this case the effect of matter
suppresses the conversion in $\nu_\mu$'s.  Since the yield of
atmospheric neutrinos is bigger than that of atmospheric
anti-neutrinos, clearly the matter effect suppresses the overall
conversion probability. Therefore we need in this case a larger value
of the vacuum mixing angle, as can be seen by comparing the left and
right lower panels in Fig.~\ref{mutausk1}.

Notice that in all channels where matter effects play a role (all,
except the upper left panel) the range of acceptable $\Delta m^2$ is
shifted towards larger values, when compared with the $\nu_\mu \to
\nu_\tau$ case. This follows from looking at the relation between 
mixing {\sl in vacuo} and in matter \cite{msw}. In fact, whenever the
magnitude of the matter potential is much larger than the difference
between the two energy eigenvalues in vacuum, i.e., when $V_\mu \gg
\Delta m^2/E$, there is a suppression of the mixing inside the Earth,
irrespective of the sign of the matter potential and/or $\Delta m^2$.
As a result, there is a lower cut in the allowed $\Delta m^2$ value,
and it lies higher than what is obtained in the data fit for the
$\nu_\mu \to \nu_\tau$ channel.  For example, let us consider the
cases of $\nu_\mu \to \nu_\tau$ and $\nu_\mu \to \nu_s$ for $\Delta
m^2>0$.  One can see comparing the upper left and lower right panels
that the values of $\Delta m^2$ for $\nu_{\mu} \to \nu_s$ channel for
the sub-GeV sample at 90\% CL are in the range from $2.0 \times
10^{-4} $eV$^2$ to $7.5 \times 10^{-3} $eV$^2$, whereas for the
$\nu_{\mu} \to \nu_\tau$ channel they are in the range from $5.1
\times 10^{-5} $eV$^2$ to $6.3 \times 10^{-3} $eV$^2$.

\begin{figure}
\centerline{\protect\hbox{\psfig{file=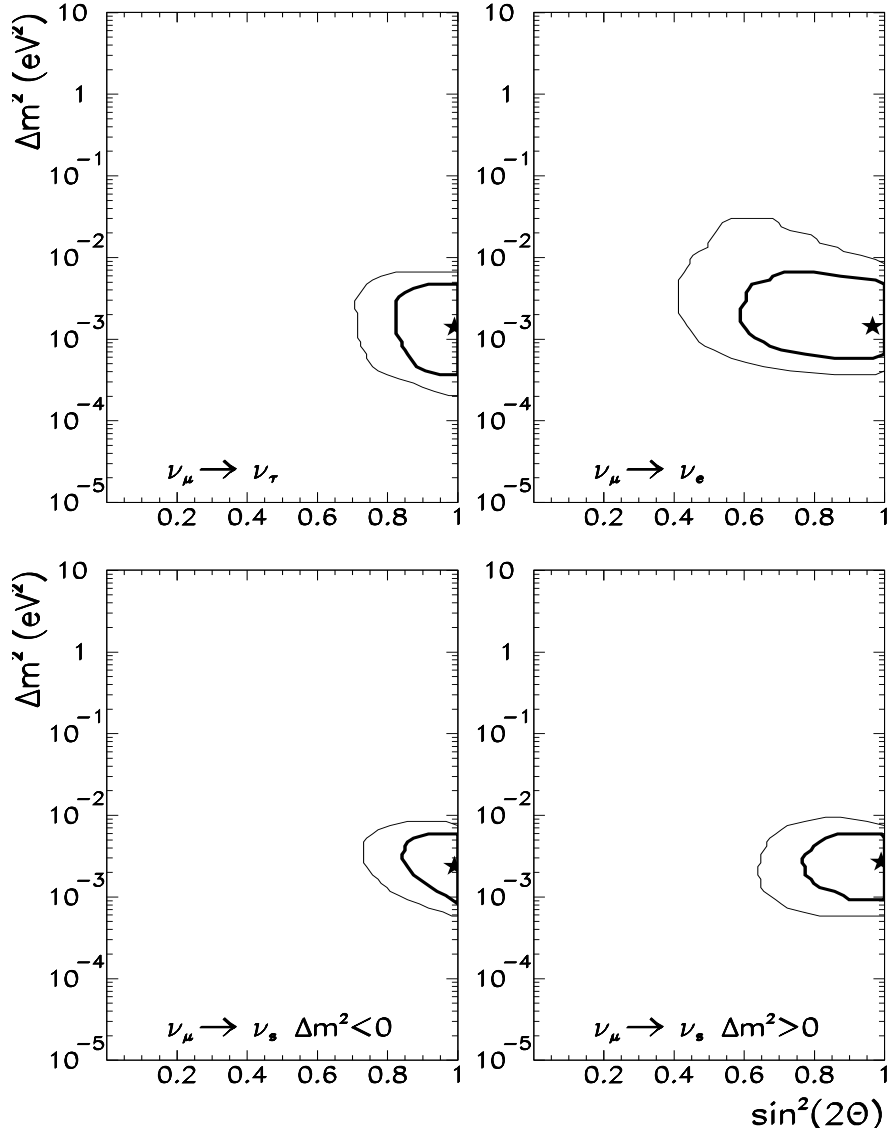,width=0.8\textwidth,height=0.7\textheight}}}
\vglue 1cm
\caption{
Allowed oscillation parameters of Super-Kamiokande data combined   
at 90 (thick solid line) and 99 \% CL  (thin solid line) for 
the different oscillation channels as labeled in the figure.
In each panel the best fit point is marked by a star.}
\label{mutausk3} 
\end{figure}

It is also interesting to analyse the effect of combining the
Super-Kamiokande sub-GeV and multi-GeV atmospheric neutrino data,
given in separate in Fig.~\ref{mutausk1}. The results corresponding to
this case are presented in Fig.\ref{mutausk3}, which also indicates
the best fit points. The quality of these fits can be better
appreciated from Tables~\ref{tab:data1} and ~\ref{tab:data2}.
Comparing the values of $\chi^2_{min}$ obtained in the present work
with those of ref.~\cite{ourwork} we see that the allowed region is
relatively stable with respect to the increased Super-Kamiokande
statistics.  However, in contrast to the case for 325.8 days, now the
$\nu_{\mu} \to \nu_\tau$ channel is as good as the $\nu_{\mu} \to
\nu_e$, when only the sub-GeV sample is included, with a clear 
Super-Kamiokande preference for the $\nu_{\mu} \to \nu_\tau$
channel. As before, the combined sub-GeV and multi-GeV data prefers
the $\nu_{\mu} \to \nu_X$, where $X=\tau$ or {\sl sterile}, over the
$\nu_{\mu} \to \nu_e$ solution.

\begin{figure}
\centerline{\protect\hbox{\psfig{file=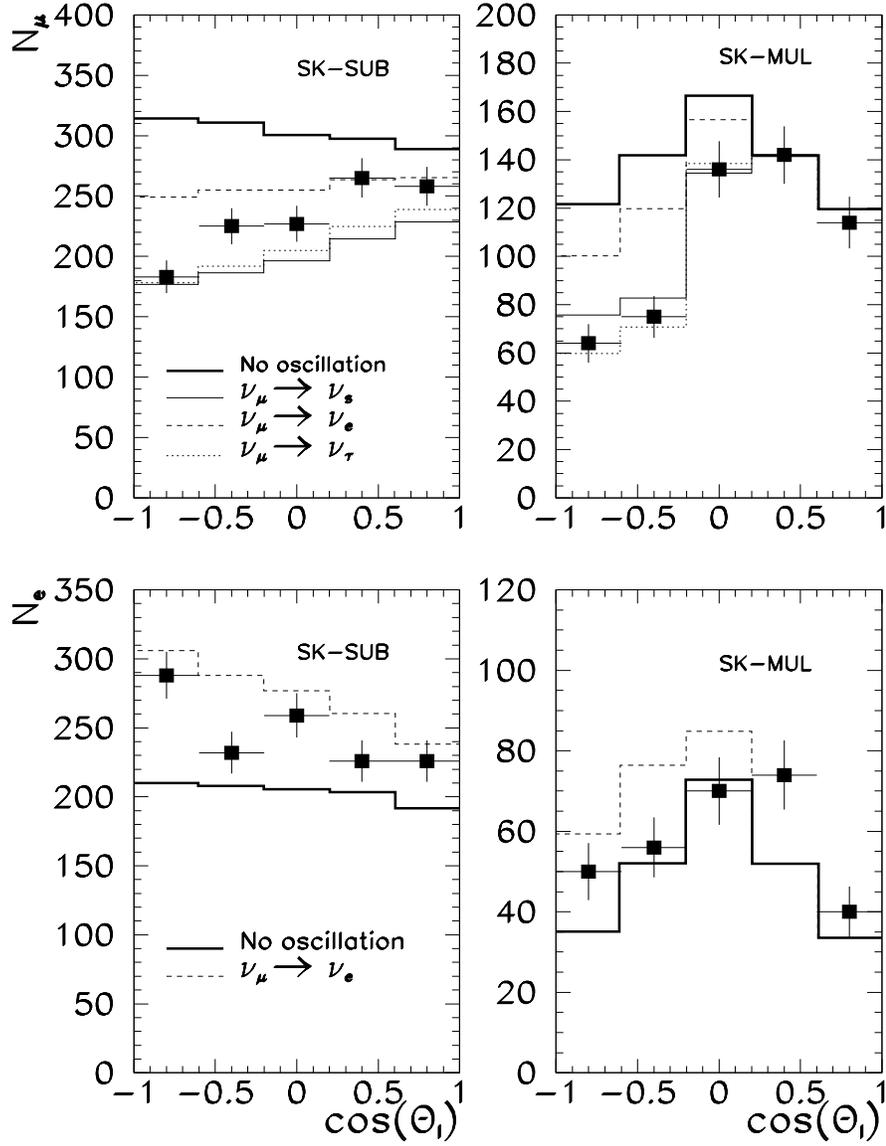,width=0.8\textwidth,height=0.7\textheight}}}
\vglue 1cm
\caption{Angular distribution for Super-Kamiokande electron-like and muon-
like sub-GeV and multi-GeV events, together with our prediction in the
absence of oscillation (thick solid line) as well as the predictions
for the best fit points in each oscillation channel: $\nu_\mu \to
\nu_s$ (thin solid line), $\nu_\mu \to \nu_e$ (dashed line) and
$\nu_\mu \to \nu_\tau$ (dotted line).  The errors displayed in the
experimental points is only statistical.}
\label{ang_mu}  
\end{figure}

The zenith angle distribution changes in the presence of oscillations
and it is different for each channel. The main information supportive
of the oscillation hypothesis comes from the Multi-GeV muon data of
Super-Kamiokande, shown in the upper right panel of Fig.~\ref{ang_mu}.
One can see that the zenith angle distribution of the electron data is
roughly consistent with the no-oscillation hypothesis. To conclude
this section we now turn to the predicted zenith angle distributions
for the various oscillation channels. As an example we take the case
of the Super-Kamiokande experiment and compare separately the sub-GeV
and Multi-GeV data with what is predicted in the case of
no-oscillation (thick solid histogram) and in all oscillation channels
for the corresponding best fit points obtained for the {\sl combined}
sub and multi-GeV data analysis performed above (all other
histograms). This is shown in Fig.~\ref{ang_mu}.  In the upper left
(right) panel we show the Super-Kamiokande sub-GeV (multi-GeV) for the
muon data and for zenith distributions of the muons expected for the
three channels $\nu_\mu \to \nu_\tau$,$\nu_\mu \to \nu_e$ and $\nu_\mu
\to \nu_s$. Notice that since the best fit point for $\nu_\mu \to
\nu_s$ occurs at $\sin(2\theta)=1$, the corresponding distributions
are independent of the sign of $\Delta m^2$.  In the lower left
(right) panel of Fig.~\ref{ang_mu} we give the same information for
the electron data. In the case of oscillations between two neutrino
species the only relevant channel to compare to in this case is the
$\nu_\mu \to
\nu_e$ channel.

It is worthwhile to see why the $\nu_\mu \to \nu_e$ channel is bad for
the Super-Kamiokande Multi-GeV data by looking at the upper right
panel in Fig.~\ref{ang_mu}. Clearly the zenith distribution predicted
in the no oscillation case is symmetrical in the zenith angle very
much in disagreement with the data. In the presence of $\nu_\mu \to
\nu_e$ oscillations the asymmetry in the distribution is much smaller than
in the $\nu_\mu \to \nu_\tau$ or $\nu_\mu \to \nu_s$ channels, as seen
from the figure.

\section{Atmospheric versus Accelerator and Reactor Experiments}

\begin{figure}
\centerline{\protect\hbox{\psfig{file=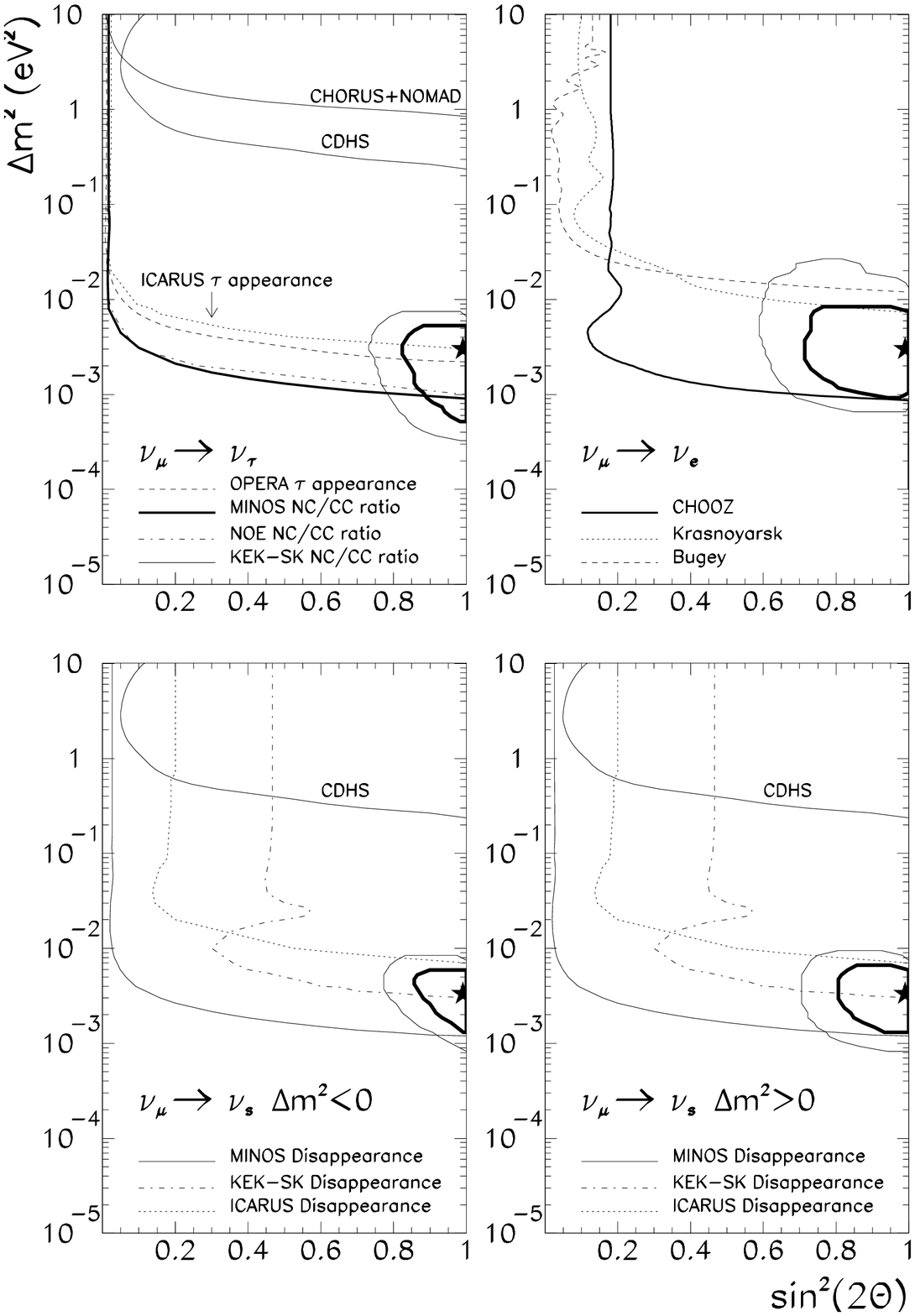,width=0.8\textwidth,height=0.65\textheight}}}
\vglue 1cm
\caption{
Allowed oscillation parameters for all experiments combined at 90
(thick solid line) and 99 \% CL (thin solid line) for each oscillation
channel as labeled in the figure.  We also display the expected
sensitivity of the future long-baseline experiments in each channel:
MINOS, KEK-SK, NOE, ICARUS and OPERA, as well as the present constraints 
of accelerator and reactor experiments: CHOOZ, Bugey and Krasnoyarsk for 
the $\nu_\mu \to \nu_e$ channel, CDHSW and CHORUS (NOMAD) for 
$\nu_\mu \to \nu_x$, 
where $x=\tau$ or sterile.  The best fit point is marked with a star.}
\label{mutausk4} 
\end{figure}

We now turn to the comparison of the information obtained from the
analysis of the atmospheric neutrino data presented above with the
results from reactor and accelerator experiments as well as the
sensitivities of future experiments. For this purpose we present the
results obtained by combining all the experimental atmospheric
neutrino data from various experiments~ \cite{atmexp+,atmexp-}. In
Fig.~\ref{mutausk4} we present the combined information obtained from
our analysis of all atmospheric neutrino data involving
vertex-contained events and compare it with the constraints from
reactor experiments like Krasnoyarsk~\cite{krasnoyarsk},
Bugey~\cite{bugey} and CHOOZ~\cite{chooz}, and the accelerator
experiments such as CDHSW \cite{CDHSW}, CHORUS and NOMAD
~\cite{dilella}. We also include in the same figure the sensitivities
that should be attained at the future long-baseline experiments now
under discussion.

The upper-left, upper-right, lower-left and lower-right panels of
Fig.~\ref{mutausk4} show respectively the global fit of all
atmospheric neutrino data for the $\nu_\mu \to \nu_\tau$, $\nu_\mu \to
\nu_e$, $\nu_\mu \to \nu_s$ with $\Delta m^2<0$ and $\nu_\mu \to
\nu_s$ with $\Delta m^2>0$. In each of these panels the star denote
the best fit point of all combined data.

The first important point is that from the upper-right panel of
Fig.~\ref{mutausk4} one sees that the CHOOZ reactor ~\cite{chooz} data
already exclude completely the allowed region for the $\nu_{\mu} \to
\nu_e$ channel when  all experiments are  combined at 90\% CL. The situation
is different if only the combined sub-GeV and multi-GeV
Super-Kamiokande are included. In such case the region obtained
(upper-right panel of Fig.~\ref{mutausk3}) is not completely excluded
by CHOOZ at 90\% CL.  Present accelerator experiments are not very
sensitive to low $\Delta m^2$ due to their short baseline. As a
result, for all channels other than $\nu_{\mu} \to \nu_e$ the present
limits on neutrino oscillation parameters from CDHSW \cite{CDHSW},
CHORUS and NOMAD \cite{dilella} are fully consistent with the region
indicated by the atmospheric neutrino analysis. Future long baseline
(LBL) experiments have been advocated as a way to independently check
the ANA.  Using different tests such long-baseline experiments now
planned at KEK (K2K) \cite{chiaki}, Fermilab (MINOS) \cite{minos} and
CERN ( ICARUS \cite{icarus}, NOE \cite{noe}, and OPERA \cite{opera}) 
would test the pattern
of neutrino oscillations well beyond the reach of present
experiments. These tests are the following:
\begin{enumerate} 
\item $\tau$ appearance searches
\item The $NC/CC$ ratio, $\Frac{(NC/CC)_{near}}{(NC/CC)_{far}}$. 
This is the most sensitive test, namely the ratio of the ratios
between the total neutral current events over charged current events
in a near detector over a far detector.  For example the curves
labelled {\sl NC/CC ratio} at the upper-left panel of
Fig.~\ref{mutausk4} delimit the sensitivity regions of these
experiments with respect to this test.
\item The muon disappearance, $CC_{near}/CC_{far}$. 
This test is based on the comparison between the number of charged
current interactions in a near detector and those measured in the far
detector.  For example, the curves labelled {\sl Disappearance } at the
lower panel of Fig.~\ref{mutausk4} delimit the sensitivity regions of
the relevant experiments with respect to this test.
\end{enumerate}
The second test can potentially discriminate between the active and sterile
channels, i.e. $\nu_\mu \to \nu_\tau$ and $\nu_\mu \to \nu_s$. However
it cannot discriminate between $\nu_\mu \to \nu_s$ and the
no-oscillation hypothesis.  In contrast, the last test can probe the
oscillation hypothesis itself.

Notice that the sensitivity curves corresponding to the disappearance
test labelled as {\sl KEK-SK Disappearance } at the lower panels of
Fig.~\ref{mutausk4} are the same for the $\nu_\mu \to \nu_\tau$ and
the sterile channel since the average energy of KEK-SK is too low to
produce a tau-lepton in the far detector ~\footnote{The KEK-SK
Collaboration is making a proposal of an upgrade aimed at seeing the
tau's~\cite{chiaki1}.}. In contrast the MINOS experiment has a higher
average initial neutrino energy and it can see the tau's. Although in
this case the exclusion curves corresponding to the disappearance test
are in principle different for the different oscillation channels, in
practice, however, the sensitivity plot is dominated by the systematic
error.  As a result discriminating between $\nu_\mu \to
\nu_{\tau}$ and $\nu_\mu \to \nu_s$ would be unlikely with the
Disappearance test\cite{goodman}.

In summary we find that, unfortunately, the regions of oscillation
parameters obtained from the analysis of the atmospheric neutrino data
on vertex-contained events cannot be fully tested by the LBL
experiments, when the Super-Kamiokande data are included in the fit
for the $\nu_\mu \to
\nu_\tau$ channel as can be seen clearly from the upper-left panel
of Fig.~\ref{mutausk4}. 
One might expect that, due to the upward shift
of the $\Delta m^2$ indicated by the fit for the sterile case, it
would be possible to completely cover the corresponding region of
oscillation parameters. This is the case for the MINOS 
disappearance test. But in general since only the disappearance 
test can discriminate against the no-oscillation hypothesis, and this
test is intrinsically weaker due to systematics, we find
that also for the sterile case most of the LBL experiments can not completely
probe the region of oscillation parameters indicated by the
atmospheric neutrino analysis. This is so irrespective of the sign of
$\Delta m^2$: the lower-left panel in Fig.~\ref{mutausk4} shows the
$\nu_\mu \to \nu_s$ channel with $\Delta m^2<0$ while the $\nu_\mu \to
\nu_s$ case with $\Delta m^2>0$ is shown in the lower-right panel.

\section{Discussion and Conclusions}
\label{conclu}

In this paper we have compared the relative quality of the
active-active and active-sterile channels as potential explanations of
the ANA using the recent 535 days sample of Super-Kamiokande as well
all available atmospheric neutrino data in the sub GeV and Multi GeV
range (vertex-contained events).  Using the most recent theoretical
atmospheric neutrino fluxes we have determined the allowed regions of
oscillation parameters for $\nu_\mu \to \nu_X$ for all possible
channels $X=e,\tau,s$. We find that the $\Delta m^2$ values for the
active-sterile oscillations (both for positive and negative $\Delta
m^2$) are higher than for the $\nu_\mu \to \nu_\tau$ case. Moreover
the increased Super-Kamiokande sample slightly favours $\nu_\mu \to
\nu_\tau$ oscillations in comparison to $\nu_\mu \to \nu_s$, and
disfavours $\nu_\mu \to \nu_e$ more strongly.  In the {\sl global} fit
of all experiments (vertex-contained events) we find a slight
preference for the $\nu_\mu \to \nu_\tau$ channel over the other
channels, including $\nu_\mu \to \nu_e$.  Insofar as the $\nu_\mu \to
\nu_e$ channel is concerned, there is no great change in the presently
allowed region of all combined experiments with respect to the earlier
situation: it is completely excluded by the CHOOZ experiment at the 90
\%CL. What is new is that the $\nu_\mu \to \nu_e$ fit for 535 days
is now worse than with the smaller sample and slightly disfavoured when
in the global fit of atmospheric data {\sl alone} and more strongly
disfavoured if Super-Kamiokande data alone are combined.

We also give the zenith angle distributions predicted for the best fit
points for each of the possible oscillation channel.  Using the zenith
angle distribution expected for multi-GeV Super-Kamiokande data in the
presence of oscillation we compare the relative goodness of the three
possible oscillation channels. This allows one to understand clearly
why the $\nu_\mu \to \nu_\tau$ and $\nu_\mu \to \nu_s$ channels are
much better than the $\nu_\mu \to \nu_e$ channel which is in any case
ruled out by CHOOZ. The main support for the \nm to \nt oscillation
hypothesis comes from the Super-Kamiokande Multi-GeV muon data.

Finally we compare our determinations of the atmospheric neutrino
oscillation parameters with the expected sensitivities of future
long-baseline experiments such as K2K, MINOS, ICARUS, NOE and OPERA. 
We have found that, unfortunately, the regions of oscillation parameters
obtained from the analysis of the atmospheric neutrino data on
vertex-contained events cannot be fully probed by most of the LBL
experiments.  Despite the $\Delta m^2$ values indicated by the
atmospheric data for the sterile case are higher than for the $\nu_\mu
\to \nu_\tau$ channel, the need to rely on the disappearance test
makes it difficult to completely test the region of atmospheric
neutrino oscillation parameters at the presently proposed LBL
experiments.  From this point of view a re-design of such experiments
would be desirable, as recently suggested by NOE\cite{noe1}.

\section*{Acknowledgements}

This work is a follow-up of the paper in ref.~\cite{ourwork} done in
collaboration with Todor Stanev, whose atmospheric neutrino fluxes we
adopt.
We thank Maury Goodmann, Takaaki Kajita, Ed Kearns and Osamu Yasuda
for useful discussion and correspondence.
It was supported by DGICYT under grant PB95-1077, by CICYT under grant
AEN96--1718, and by the TMR network grant ERBFMRXCT960090 of the
European Union. H. Nunokawa and O. Peres were supported by FAPESP
grants.

\vglue 1cm

\newpage 
\begin{table}[h]
\begin{center}
\begin{tabular}{||l|l|l|l|l|||}
Experiment &  & $\nu_{\mu} \to  \nu_\tau$ &  
                $\nu_{\mu} \to  \nu_s$   &
                $\nu_{\mu} \to  \nu_e$  
\\\hline
Super-Kam   & $\chi^2_{min}$ 
                           & $  6.1 $ & $  7.7 $ & $ 7.3$
\\
sub-GeV                   & $ \Delta m^2 $ ( $10^{-3} $eV$^2$ )
                           & $ 0.11 $    & $1.6$   & $1.2$
\\
                           & $\sin^2 2\theta$ 
                           &  $1.0$   & $1.0$   & $0.97$ 
\\\hline

Super-Kam  & $\chi^2_{min}$ 
                           & $  6.3$ & $  7.9 $ & $ 10.8$
\\
multi-GeV                 & $ \Delta m^2 $ ( $10^{-3} $eV$^2$ )
                           & $1.4$    & $3.5$   & $25.4$
\\
                           & $\sin^2 2\theta$ 
                           &  $0.97$   & $1.0$   & $0.74$
\\\hline 
\end{tabular}
\end{center}
\vglue 1cm
\caption[Tab]{Minimum value of $\chi^2$ and the best fit point for 
Super-Kamiokande sub-GeV and Multi-Gev data and for each oscillation
channel. 
Notice that for 
$\nu_\mu\rightarrow \nu_s$ the minimum $\chi^2$ is practically independent
of the sign of $\Delta m^2$ as the minimum is located at maximum mixing 
angle.}
\label{tab:data1}
\end{table}

\begin{table}
\begin{center}
\begin{tabular}{||l|l|l|l|l|l|||}
Experiment &  & $\nu_{\mu} \to  \nu_\tau$ &  
                $\nu_{\mu} \to  \nu_s$   &
                $\nu_{\mu} \to  \nu_e$  
\\\hline
Super-Kam   & $\chi^2_{min}$ 
                           &$ 13.7$ & $ 16.5 $ & $21.9$
\\
    Combined              & $ \Delta m^2 $ ( $10^{-3} $eV$^2$ ) 
                           & $1.4$    & $2.2$   & $1.4$
\\
                           & $\sin^2 2\theta$ 
                           &  $1.0$   & $1.0$   & $0.96$ 
\\\hline
All experiments    & $\chi^2_{min}$ 
                           &$ 49.5$ & $ 50.8 $ & $50.8$
\\
  Combined               & $ \Delta m^2 $ ( $10^{-3} $eV$^2$ ) 
                           & $3.0$    & $3.3$  & $3.0$
\\
                           & $\sin^2 2\theta$ 
                           &  $1.0$   & $1.0$   & $0.99$ 
\\\hline
\end{tabular}
\end{center}
\vglue 1cm
\caption[Tab]{Minimum value of $\chi^2$ from for the combined data 
of Super-Kamiokande only (upper part of table) and for all experiments
(lower part of table) and the best fit point for each type of
oscillation channels. }
\label{tab:data2}
\end{table}
 

\begin{references}
\bibitem{noe1} 
After the completion of ref.~\cite{ourwork}, the NOE Collabor\-ation
has proposed a redesign of the experiment. We show the updated figure
from ~\cite{noe}.
\end{references}
\end{document}